\documentstyle[preprint,aps,epsfig]{revtex}

\begin{document}

\draft

\title{Spatial Correlations in Chaotic Eigenfunctions}

\author{Mark Srednicki\footnote{E--mail: \tt mark@tpau.physics.ucsb.edu} }

\address{Department of Physics, University of California,
         Santa Barbara, CA 93106 }

\maketitle

\tighten

\begin{abstract}
\normalsize{
At short distances, energy eigenfunctions of chaotic systems have 
spatial correlations that are well described by assuming a microcanonical
density in phase space for the corresponding Wigner function.  However, 
this is not correct on large scales.  The correct correlation function 
is in turn needed to get the correct formula for the root-mean-square value
of the off-diagonal matrix elements of simple observables,
and for the fluctuations in the diagonal elements.
\vskip1in
\centerline{\em Invited talk given at the NATO Advanced Study Institute,}
\centerline{\em Supersymmetry and Trace Formulae,}
\centerline{\em Isaac Newton Institute, Cambridge, UK}
\centerline{\em 8--19 September 1997}
}
\end{abstract}


My focus in this talk will be on the statistical properties of 
the quantum energy eigenfunctions in classically chaotic systems.
Twenty years ago, Berry \cite{berry77}
conjectured that these eigenfunctions could be treated as
gaussian random variables, and that the spatial correlations
would be those following from taking the expected value of the
Wigner density for each eigenfunction to be microcanonical.
More specifically, the probability that 
the actual eigenfunction $\psi_\alpha({\bf q})$ for energy
$E=E_\alpha$ (the corresponding energy eigenvalue)
is between $\psi({\bf q})$ and $\psi({\bf q}) + d\psi({\bf q})$ 
for all coordinate points $\bf q$ is given by 
\begin{equation}
P(\psi|E) \propto \exp
            \left[-{\beta\over2}\int d^f\! q\int d^f\! q'\ 
                   \psi^*({\bf q})K({\bf q},{\bf q}'|E)\psi({\bf q}')
            \right]{\cal D}\psi \; .
\label{prob}
\end{equation}
If the system is time-reversal invariant,
the eigenfunctions are real, $\beta=1$,
and the measure is ${\cal D}\psi = \prod_{\bf q} d\psi({\bf q})$;
if it is not, the eigenfunctions are complex, $\beta=2$, and the measure is
${\cal D}\psi = \prod_{\bf q} d\mathop{\rm Re}\psi({\bf q}) \,
                              d\mathop{\rm Im}\psi({\bf q})$.
In either case, the kernel $K({\bf q},{\bf q}'|E)$
is the functional inverse of the two-point correlation function
\begin{equation}
C({\bf q}',{\bf q}|E) 
\equiv \langle \psi({\bf q}')\psi^*({\bf q}) \rangle \; ,
\label{cdef}
\end{equation}
where the angle brackets denote averaging over $P(\psi|E)$.
Taking the Wigner density to be microcanonical results in
\begin{equation}
C({\bf q}',{\bf q}|E) = {1\over{\bar\rho}(E)}
                           \int {d^f\! p \over (2\pi\hbar)^f}\; 
                                e^{i{\bf p}\cdot({\bf q}'-{\bf q})/\hbar}
                                \delta(E-H_W({\bf p},{\bf \bar q})) \; ,
\label{berry}
\end{equation}
where ${\bf \bar q} = {1\over2}({\bf q}+{\bf q}')$ is the midpoint, 
$H_W({\bf p},{\bf q})$ is the classical hamiltonian
(more specifically, it is the Weyl symbol of the hamiltonian operator),
and ${\bar\rho}(E)$ is the semiclassical density of states,
\begin{equation}
{\bar\rho}(E) = \int {d^f\! p \, d^f\! q \over (2\pi\hbar)^f}\; 
                     \delta(E-H_W({\bf p},{\bf q})) \; .
\label{rhobar}
\end{equation}

Berry's conjecture for $C({\bf q}',{\bf q}|E)$
can be tested numerically in two-dimensional billiard
systems; in this case, $H={\bf p}^2/2m$, and we have
$C({\bf q}',{\bf q}|E) = J_0(k|{\bf q}'-{\bf q}|)$,
where $E=\hbar^2 k^2/2m$ and $J_0(x)$ is an ordinary Bessel function.
An ``experimental'' correlation function 
$C_{\rm exp}({\bf s},D|E)$ can be defined in terms of a particular 
energy eigenfunction $\psi_\alpha({\bf q})$ via
\begin{equation}
C_{\rm exp}({\bf s},D|E)
\equiv {A_B\over A_D}\int_D d^2q \;
\psi_\alpha^*({\bf q}+{\textstyle{1\over2}}{\bf s}) 
\psi_\alpha({\bf q}-{\textstyle{1\over2}}{\bf s}) \;,
\label{cexp}
\end{equation}
where $E=E_\alpha$, $A_B$ is the area of the billiard, 
and $A_D$ is the area of a domain~$D$
over which the central point ${\bf q}$ is averaged.
Early computations \cite{mk,as} of $C_{\rm exp}({\bf s},D|E)$
found that it typically resembled its predicted value $J_0(k|{\bf s}|)$,
but with large fluctuations.  Later it was realized \cite{sred94}
that these fluctuations were in fact entirely consistent with the 
expected gaussian fluctuations \cite{ogh} of $\psi_\alpha({\bf q})$.
Recent detailed computations \cite{lirob,ss96} leave
little doubt that (at asymptotically high energies) the eigenfunctions of 
chaotic billiards are very well described by Berry's conjecture.

Berry's conjecture also follows from other approaches to quantum chaos.
The random-matrix analogy \cite{bgs,bohigas}
suggests that the energy eigenstates have a Porter-Thomas
distribution of amplitudes in a suitably chosen basis.
For billiards, the natural choice is the momentum basis, and from this
starting point one can derive eqs.~(\ref{prob}) and 
(\ref{berry}) \cite{alhas}.  If one adds a white-noise,
spatially uncorrelated random potential to the system, and takes
the limit where the mean-free path is much larger than the system size,
then eq.~(\ref{prob}) can be derived via the supersymmetric
sigma-model technique \cite{pei,paei,prig1,prig2,sred96,blamir}.

However, we cannot expect eq.~(\ref{berry}) to be valid
when the separation between ${\bf q}$ and ${\bf q}'$ becomes
large compared to distance scales which are important classically.
This is because
the hamiltonian evaluated at just the midpoint ${\bf \bar q}$
does not contain enough information about the classical landscape
between ${\bf q}$ and ${\bf q}'$.
To find the correct formula for $C({\bf q}',{\bf q}|E)$ 
when $|{\bf q}'-{\bf q}|$ is large \cite{hs1},
we first consider the theory with an added random potential.
In this case an expression for 
$C({\bf q}',{\bf q}|E)$
can be derived explicitly \cite{paei}:
\begin{equation}
C({\bf q}',{\bf q}|E) = {1\over 2\pi i {\bar\rho}(E)} 
                           \left[  {\overline G}({\bf q},{\bf q}'|E)^* 
                                 - {\overline G}({\bf q}',{\bf q}|E)   
                           \right] \; .
\label{cgbar}
\end{equation}
Here $G({\bf q}',{\bf q}|E)$ is the energy Green's function,
\begin{equation}
G({\bf q}',{\bf q}|E) =
\sum_\alpha {\psi_\alpha({\bf q}')\psi_\alpha^*({\bf q}) 
             \over E-E_\alpha+i0^+}  \; ,
\label{green}
\end{equation}
$\rho(E)$ is the density of states,
\begin{equation}
\rho(E) = {1\over 2\pi i} \int d^f\! q \left[  G({\bf q},{\bf q}|E)^* 
                                               - G({\bf q},{\bf q}|E)   
                                         \right] \; ,  
\label{rho}
\end{equation}
and the bar stands for averaging over the random potential.
We now make the assumption that, for a chaotic system, 
eq.~(\ref{cgbar}) continues to hold if we first take the limit
of small $\hbar$, and then disregard the random potential.

There are two different semiclassical approximations
which can be used to compute the Green's function in the small-$\hbar$ 
limit \cite{gutz67,bm72}.  
The first applies when the distance between ${\bf q}$ and ${\bf q}'$
is small, in the sense that the classical path of least action 
with energy $E$ which connects
${\bf q}$ to ${\bf q}'$
is well approximated by a linear function of time (in any coordinate
system).  This path then dominates the Green's function, and one finds
\begin{equation}
{\overline G}({\bf q}',{\bf q}|E) 
=\int {d^f\!p\over(2\pi\hbar)^f} \; 
      e^{i{\bf p}\cdot({\bf q}'-{\bf q})/\hbar} \;
      {1\over E-H_W({\bf p},{\bf \bar q})+i0^+} \;,
\label{greenclass}
\end{equation}
which immediately yields eq.~(\ref{berry}) for
$C({\bf q}',{\bf q}|E)$.
If, however, the classical path of least action is not well approximated
by a linear function of time, and the value of this action is much larger
than $\hbar$, then we have instead
\begin{equation}
{\overline G}({\bf q}',{\bf q}|E) = 
{1\over i\hbar(2\pi i\hbar)^{(f-1)/2}} \sum_{\rm paths}
         |D_{\rm p}|^{1/2}e^{iS_{\rm p}/\hbar-i\nu_{\rm p}\pi/2} \; .
\label{greengutz}
\end{equation}
Here the sum is over all classical paths connecting 
${\bf q}$ to ${\bf q}'$ with energy $E$, action
\begin{equation}
S_{\rm p} = \int_{\bf q}^{{\bf q}'} {\bf p}\cdot d{\bf q} \; ,
\label{s}
\end{equation}
focal point number $\nu_{\rm p}$, and fluctuation determinant
\begin{equation}
D_{\rm p} = \det\pmatrix{ {\partial^2 S_{\rm p} \over
                            \partial{\bf q}_2 \partial{\bf q}_1 } &
                           {\partial^2 S_{\rm p} \over
                            \partial E \partial{\bf q_1}        } \cr
\noalign{\medskip}
                           {\partial^2 S_{\rm p} \over
                            \partial{\bf q}_2 \partial E        } &
                           {\partial^2 S_{\rm p} \over
                            \partial E^2                        } \cr } \; .
\label{det}
\end{equation}
Eqs.~(\ref{cgbar}) and (\ref{greengutz})
give the correct formula for
$C({\bf q}',{\bf q}|E)$ when $|{\bf q}'-{\bf q}|$ is large \cite{hs1}.

This formula is most useful under circumstances where the sum over paths
is dominated by a single path of least action.  An interesting example 
is a two-dimensional billiard in a 
perpendicular magnetic field $\bf B$.  In the billiard
interior the hamiltonian is $H=({\bf p}-e{\bf A})^2/2m$, and we
will work in the gauge in which the vector potential is
${\bf A}={1\over2}{\bf B}\times{\bf q}$.
Assuming that it is not blocked,
the classical path of least action is a circular arc with length $\ell$,
related to the separation $L=|{\bf q}' -{\bf q}|$ and classical cyclotron
radius $R = (2mE)^{1/2}/|eB|$ via $\ell=2R\sin^{-1}(L/2R)$.  The action
for this path can be divided into a geometric part and a gauge-dependent part,
$S_{\rm p} = S_{\rm geom} + S_{\rm gauge}$.  
The geometric part is
\begin{eqnarray}
S_{\rm geom}  &=& \hbar k \left(\ell-{{\cal A}\over R}\right) 
\nonumber \\
\noalign{\medskip}
              &=& \hbar kL\left(1-{L^2\over 24 R^2} +\ldots\right) \; ,
\label{sgeo} 
\end{eqnarray}
where $\hbar k=(2mE)^{1/2}$,
and ${\cal A} = {1\over2}R \ell - {1\over2} R^2 \sin(\ell/R)$ 
is the area enclosed by the circular arc and the straight line connecting
${\bf q}$ to ${\bf q}'$.  
For our gauge choice, the gauge-dependent part is
\begin{equation}
S_{\rm gauge} = {\textstyle{1\over 2}} 
                  e {\bf B}\!\cdot\!({\bf q}\times{\bf q}') \; .
\label{sgauge}
\end{equation}
In any gauge, the determinant $|D_{\rm p}|$ is given by
\begin{equation}
|D_{\rm p}| = {m^2 \over \hbar kL}\left(1-{L^2\over 4 R^2}\right)^{-1/2} \; .
\label{dpb}
\end{equation}
Keeping only the contribution of this one path, and recalling that 
$\bar\rho(E)=mA_B/2\pi\hbar^2$ for a two-dimensional billiard
with area $A_B$, we find from 
eqs.~(\ref{cgbar}) and (\ref{greengutz}) that
\begin{equation}
C({\bf q}',{\bf q}|E) = A_B^{-1}\exp(iS_{\rm gauge}/\hbar)
                           {\cos(S_{\rm geom}/\hbar -\pi/4) \over
                           (\pi kL/2)^{1/2}
                           (1-L^2/4 R^2)^{1/4} } \; .
\label{cb}
\end{equation}
We need $kL\gg 1$ for this formula to hold.  Also, to avoid an integrable
region of phase space, a circle of radius $R$ must not be able to fit
inside the billiard; this implies $L<2R$.

We can compare eq.~(\ref{cb}) with the result of using Berry's formula,
eq.~(\ref{berry}); with our gauge choice we find
\begin{equation}
C({\bf q}',{\bf q}|E) = A_B^{-1}\exp(iS_{\rm gauge}/\hbar) J_0(kL) \;.
\label{cbb}
\end{equation}
For large $kL$, we can use the asymptotic form of the Bessel function to get
\begin{equation}
C({\bf q}',{\bf q}|E) = A_B^{-1}\exp(iS_{\rm gauge}/\hbar)
                           {\cos(kL-\pi/4) \over
                           (\pi kL/2)^{1/2} }     \;.
\label{cbb2}
\end{equation}
We see that Berry's formula misses the corrections due to the finite cyclotron
radius $R$ which are present in eq.~(\ref{cb}).

If we make a gauge transformation 
\begin{equation}
{\bf A}({\bf q}) \to {\bf A}({\bf q}) + \nabla\Phi({\bf q}) \; ,
\label{trans1}
\end{equation}
where $\Phi({\bf q})$ is any smooth function, then
\begin{equation}
S_{\rm gauge} \to S_{\rm gauge} + e\Phi({\bf q}') - e\Phi({\bf q}) \; ,
\label{trans2}
\end{equation}
and so eq.~(\ref{cb}) implies
\begin{equation}
C({\bf q}',{\bf q}|E) \to e^{+ie[\Phi({\bf q}')-\Phi({\bf q})]/\hbar}
                             C({\bf q}',{\bf q}|E) \; .
\label{ctr}
\end{equation}
That this is correct can be seen by recalling that a wave function
$\psi({\bf q})$ transforms as 
\begin{equation}
\psi({\bf q)} \to e^{+ie\Phi({\bf q})/\hbar}\psi({\bf q)}
\label{psitr}
\end{equation}
under (\ref{trans1}), and that $C({\bf q}',{\bf q}|E_\alpha)$
is the expected value of $\psi_\alpha({\bf q}')\psi_\alpha^*({\bf q})$.
On the other hand, eq.~(\ref{berry}) implies instead that 
\begin{equation}
C({\bf q}',{\bf q}|E) \to e^{+ie({\bf q}'-{\bf q})\cdot
                                   \nabla\Phi({\bf \bar q})/\hbar}
                             C({\bf q}',{\bf q}|E) \; ,
\label{cbtr}
\end{equation}
which again illustrates the fact that Berry's formula is valid only when
$|{\bf q}'-{\bf q}|$ is sufficiently small.

I now turn to another issue, the statistical properties of transition
matrix elements $A_{\alpha\beta} \equiv \langle\alpha|A|\beta\rangle$
in chaotic systems.
Here the operator $A$ is a smooth, $\hbar$-independent
function of the coordinates and momenta, and $|\alpha\rangle$ and 
$|\beta\rangle$ are different energy eigenstates.
We will be interested in the average value of
$|A_{\alpha\beta}|^2$ when the energies $E_\alpha$ and $E_\beta$
are each varied over a range which is small classically,
but encompasses many quantum energy levels.
For later convenience let us define
\begin{equation}
\bar E = {\textstyle{1\over2}}(E_\alpha+E_\beta)
\qquad \hbox{and} \qquad
\hbar\omega = E_\beta-E_\alpha \; .
\label{ebar}
\end{equation}
We will do our computations in the limit $\hbar\to 0$ with
$\bar E$ and $\omega$ held fixed.  The mean level spacing
near energy $\bar E$ is $1/\bar\rho(\bar E) \sim \hbar^f$,
so our condition on the ranges of $E_\alpha$ and $E_\beta$
is easy to fulfill.

One approach \cite{fp,wilk,pros}
to this problem is to relate the average value
$\langle|A_{\alpha\beta}|^2\rangle$ to the classical 
time-correlation function of the observable $A$ via
\begin{equation}
\langle|A_{\alpha\beta}|^2\rangle 
= {1\over\tau_{\rm H}}
  \int^{\tau_{\rm H}}_{-\tau_{\rm H}} dt \, 
                           e^{i\omega t}
                           \int d\mu_{\bar E} \, 
                           A_W({\bf p}_t,{\bf q}_t)A_W({\bf p},{\bf q}) \; .
\label{aab1}
\end{equation}
Here $\tau_{\rm H}=2\pi\hbar\bar\rho(\bar E)$ is the Heisenberg time,
$({\bf p}_t,{\bf q}_t)$ is the point in phase space which is reached 
classically at time $t$ when starting from $({\bf p},{\bf q})$ at time zero,
$A_W({\bf p},{\bf q})$ is the Weyl symbol of the operator $A$,
and $d\mu_E$ is the Liouville measure on the surface in phase
space with energy $E$, given by
\begin{equation}
d\mu_E = {1\over\bar\rho(E)}\;
         {d^f\! p \, d^f\! q \over (2\pi\hbar)^f}\; 
         \delta(E-H_W({\bf p},{\bf q})) \; .
\label{mue}
\end{equation}
Note that the notation, though useful later, is somewhat misleading:
$d\mu_E$ is a purely classical, $\hbar$-independent object.
We will not discuss the derivation \cite{fp,wilk,pros}
of eq.~(\ref{aab1}) here; a brief and non-rigorous review is given 
elsewhere \cite{hs2}.

Another approach \cite{sred94,eck2,ss96}
to computing $\langle|A_{\alpha\beta}|^2\rangle$ 
is to write $|A_{\alpha\beta}|^2$ in terms of the energy eigenfunctions
$\psi_\alpha({\bf q})$ and  
$\psi_\beta({\bf q})$, and then average
$|A_{\alpha\beta}|^2$ over the eigenfunction probability distribution
of eq.~(\ref{prob}).  This calculation is simplified if $A$ is a function of
${\bf q}$ only (rather than both $\bf q$ and $\bf p$), and so we
specialize to this case.  We get
\begin{equation}
\langle|A_{\alpha\beta}|^2\rangle 
= \int d^f\! q' \, d^f\! q \; C({\bf q},{\bf q}'|E_\alpha) 
                              A_W({\bf q}')
                              C({\bf q}',{\bf q}|E_\beta) 
                              A_W({\bf q}) \; .
\label{aab2}
\end{equation}
It is not at all clear that eq.~(\ref{aab2}) gives the same result
for $\langle|A_{\alpha\beta}|^2\rangle$
as eq.~(\ref{aab1}), an issue which was raised (in the context
of the diagonal matrix elements) by Austin and Wilkinson \cite{aw}.
If Berry's formula is used for $C({\bf q}',{\bf q}|E)$,
then eqs.~(\ref{aab1}) and (\ref{aab2}) do not necessarily agree,
as can be seen by working out some simple examples.
In this case, it is eq.~(\ref{aab2}) which is wrong.
The reason is that $\bf q$ and ${\bf q}'$ are independently integrated
in eq.~(\ref{aab2}), and so $|{\bf q}'-\bf q|$ is generically large.
Therefore we should use eqs.~(\ref{cgbar}) and (\ref{greengutz}) for
$C({\bf q}',{\bf q}|E)$, rather than eq.~(\ref{berry}).  
If we do, then eqs.~(\ref{aab1})
and (\ref{aab2}) give the same result for
$\langle|A_{\alpha\beta}|^2\rangle$.

To see this still requires some work.  A detailed exposition is
given elsewhere \cite{hs2}, and here we will just highlight
the key elements.  One is the diagonal approximation \cite{berry85};
after substituting eqs.~(\ref{cgbar}) and (\ref{greengutz}) into 
eq.~(\ref{aab2}), the double sum over paths is collapsed to a single sum,
since the off-diagonal terms will have rapidly oscillating
phases.  In different but related contexts \cite{berry85,eck2}, 
this requires restricting the single sum
to paths whose elapsed times are less than the Heisenberg time,
and we will assume the same is true here.
The action difference in the single sum is
\begin{eqnarray}
S_{{\rm p}_\beta} - S_{{\rm p}_\alpha}
&=& (E_\beta-E_\alpha){\partial S_{\rm p}\over\partial E}\Biggr|_{E=\bar E} 
    + \ldots
\nonumber \\
&=& \hbar \omega \tau_{\rm p} + O(\hbar^2) \;,
\label{taup}
\end{eqnarray}
where $\tau_{\rm p}$ is the elapsed time along the path.
We assume that
$\nu_{{\rm p}_\beta}=\nu_{{\rm p}_\alpha}$,
since $\nu_{\rm p}$ is a topological quantity which in general will not change
when the energy of the path is varied slightly.
Finally, we note that the determinant $D_{\rm p}$ can be written as
\begin{equation}
D_{\rm p} =   \det\pmatrix{ -{\partial{\bf p}  \over
                              \partial{\bf q}'                   } &
                             {\partial\tau     \over
                              \partial{\bf q}'                   } \cr
\noalign{\medskip}
                            -{\partial{\bf p}  \over
                              \partial E                         } &
                             {\partial\tau     \over
                              \partial E                         } \cr } \; .
\label{det2}
\end{equation}
Here ${\bf p} = -\partial S_{\rm p}/\partial{\bf q}$
is the momentum at the beginning of the path, 
and $\tau=\tau_{\rm p}$ is the elapsed time along the path.
Eq.~(\ref{det2}) shows us that $|D_{\rm p}|$ can be 
thought of \cite{gutz67}
as a jacobian for a change of variables from the final position ${\bf q}'$
and total energy $\bar E$ to the initial momentum ${\bf p}$ and elapsed time
$\tau$.  Eq.~(\ref{aab2}) already has an integral over ${\bf q}'$,
and to get one over $\bar E$ we insert
$1=\int d\bar E\,\delta(\bar E - H_W({\bf p},{\bf q}))$.
Now we can make the change of integration variables suggested by
eq.~(\ref{det2}), and we find
\begin{equation}
\langle|A_{\alpha\beta}|^2\rangle
= {1\over\pi\hbar{\bar\rho}(\bar E)^2}
  \int_0^{\tau_{\rm H}} d\tau
  \int {d^f\! p\, d^f\! q \over (2\pi\hbar)^f} 
  \sum_{\rm paths}
  \delta(\bar E - H_W({\bf p},{\bf q}))
  \cos(\omega\tau)
  A_W({\bf q}')A_W({\bf q}) \; .
\label{aab4}
\end{equation}
The sum is now over all paths which begin at $({\bf p},{\bf q})$ and have
elapsed time $\tau$.  However, there is only one such path, 
and so the sum over paths may be dropped.
Also, ${\bf q}'$ is the position at time $\tau$, and it is now more properly
denoted ${\bf q}_\tau$.  Using eq.~(\ref{mue}),
the fact that time-translation invariance implies that 
$\int d\mu_{\bar E}\, A_W({\bf q}_\tau)A_W({\bf q})$ is an even function
of $\tau$ (even if the system is not time-reversal invariant), 
and $\tau_{\rm H} = 2\pi\hbar\bar\rho(\bar E)$,
we see that eq.~(\ref{aab4}) can be rewritten as
\begin{equation}
\langle|A_{\alpha\beta}|^2\rangle
=   {1\over\tau_{\rm H}}
    \int_{-\tau_{\rm H}}^{+\tau_{\rm H}} d\tau \,
    e^{i\omega\tau} 
    \int d\mu_{\bar E}\,
    A_W({\bf q}_\tau)A_W({\bf q}) \; ,
\label{aab6}
\end{equation}
which is equivalent to eq.~(\ref{aab1}).

Another quantity of interest is the size of the fluctuations in the
diagonal matrix elements $A_{\alpha\alpha}$.
If we first shift $A$ (if necessary) so that
$\langle A_{\alpha\alpha}\rangle = \int d\mu_{E_\alpha} \, A_W({\bf q}) = 0$,
then the object we wish to evaluate is $\langle |A_{\alpha\alpha}|^2\rangle$.
This has been done previously \cite{wilk2,eck1,eck2}
by making use of the trace formula \cite{gutz67,gutz2,eck3}
and properties of periodic orbits.
Here we will instead compute $\langle |A_{\alpha\alpha}|^2\rangle$
by averaging over the probability distribution for 
energy eigenfunctions \cite{sred94,eck2,ss96,hs2}.
In the case of a system which is not invariant 
under time reversal, the energy eigenfunctions are generically complex, 
and the relevant formula is \cite{sred96}
\begin{equation}
\langle\psi^*_1\psi_2\psi^*_3\psi_4\rangle
=
\langle\psi^*_1\psi_2\rangle
\langle\psi^*_3\psi_4\rangle
+
\langle\psi^*_1\psi_4\rangle
\langle\psi^*_3\psi_2\rangle
\; ,
\label{wick1}
\end{equation}
where $\psi_i=\psi_\alpha({\bf q}_i)$.
If the system is invariant under time reversal,
the energy eigenfunctions are real, and we have instead \cite{sred96}
\begin{equation}
\langle\psi_1\psi_2\psi_3\psi_4\rangle
=
\langle\psi_1\psi_2\rangle
\langle\psi_3\psi_4\rangle
+
\langle\psi_1\psi_4\rangle
\langle\psi_2\psi_3\rangle
+
\langle\psi_1\psi_3\rangle
\langle\psi_2\psi_4\rangle
\; .
\label{wick2}
\end{equation}
Combined with the previous results for $\langle|A_{\alpha\beta}|^2\rangle$,
we find that
\begin{equation}
\langle |A_{\alpha\alpha}|^2\rangle
= {2/\beta \over \tau_{\rm H}}
  \int^{\tau_{\rm H}}_{-\tau_{\rm H}} d\tau
                        \int d\mu_{E_\alpha} \, 
                        A_W({\bf q}_\tau)A_W({\bf q}) \; .
\label{aaa}
\end{equation}
Here $\beta=1$ for a system which is invariant under time reversal,
and $\beta=2$ for a system which is not.  Eq.~(\ref{aaa}) is in agreement
with the earlier results \cite{wilk2,eck1,eck2}.

\begin{acknowledgments}

This work was supported in part by NSF Grant PHY--97--22022.

\end{acknowledgments}

\end{document}